\documentclass[11pt]{article}
\usepackage{ifthen}
\usepackage{natbib}
\input{epsf}

\def\ltsima{$\; \buildrel < \over \sim \;$}
\def\simlt{\lower.5ex\hbox{\ltsima}}
\def\gtsima{$\; \buildrel > \over \sim \;$}
\def\simgt{\lower.5ex\hbox{\gtsima}}
\def\kpc{{\rm\,kpc}}

\def\s{\ifmmode \widetilde \else \~\fi}
\def\={\overline}

\def\spose#1{\hbox to 0pt{#1\hss}}

\def\lta{\mathrel{\spose{\lower 3pt\hbox{$\mathchar"218$}}
     \raise 2.0pt\hbox{$\mathchar"13C$}}}
\def\gta{\mathrel{\spose{\lower 3pt\hbox{$\mathchar"218$}}
     \raise 2.0pt\hbox{$\mathchar"13E$}}}
\def\Dt{\spose{\raise 1.5ex\hbox{\hskip3pt$\mathchar"201$}}}    % upper case
\def\dt{\spose{\raise 1.0ex\hbox{\hskip2pt$\mathchar"201$}}}    % lower case

\def\dotsfill{\leaders\hbox to 1em{\hss.\hss}\hfill}

\newcommand{\Line}[0]{%
  \rule{0cm}{0cm}\\\hrule\rule{0cm}{0cm}%
}

\begin{document}

\begin{flushright}
Final version: 27 April 2001 \\
\vspace{-1.0cm}
\Line
\end{flushright}
\vspace{0.3cm}
{\bf \large A giant stream of metal-rich stars in the halo of the galaxy M31}

\begin{flushleft}

{Rodrigo Ibata$^{\spadesuit}$,
Michael Irwin$^{\dagger}$,
Geraint Lewis$^{\ddagger}$,
Annette Ferguson$^*$
and Nial Tanvir$^{\parallel}$  }

\bigskip
{\small
{$^{\spadesuit}$ Observatoire de Strasbourg, 11, rue de l'Universite, F-67000
Strasbourg, France}

{$^\dagger$ Institute of Astronomy, Madingley Road, Cambridge, CB3 0HA, UK
      }

{$^\ddagger$ Anglo-Australian Observatory, PO Box 296, Epping,
NSW 1710, Australia}

{$^*$ Kapteyn Astronomical Institute, Postbus 800, 9700 AV Groningen, The Netherlands}

{$^\parallel$ Physical Sciences, Univ. of Hertfordshire,
Hatfield, AL10 9AB, UK}
}
\end{flushleft}
\vskip -1.25cm
\Line
\vskip 0.2cm

% Start Here:

{\bf Recent observations have revealed streams of gas and stars in the
halo of the Milky Way$^{1-3}$ that are the debris from interactions
between our Galaxy and some of its dwarf companion galaxies; the
Sagittarius dwarf galaxy and the Magellanic clouds. Analysis of the
material has shown that much of the halo is made up of cannibalized
satellite galaxies$^{2,4}$, and that dark matter is distributed nearly
spherically in the Milky Way. It remains unclear, however, whether
cannibalized substructures are as common in the haloes of galaxies as
predicted by galaxy-formation theory$^5$. Here we report the discovery
of a giant stream of metal-rich stars within the halo of the nearest
large galaxy, M31 (the Andromeda galaxy). The source of this stream
could be the dwarf galaxies M32 and NGC205, which are close companions
of M31 and which may have lost a substantial number of stars owing to
tidal interactions. The results demonstrate that the epoch of galaxy
building still continues, albeit at a modest rate, and that tidal
streams may be a generic feature of galaxy haloes.  }

Within the framework of hierarchical structure formation, large spiral
galaxies like the Milky Way or Andromeda arose from the merger of many
small galaxies and protogalaxies$^6$. Later in their evolution, spiral
galaxies become the dominant component in such mergers, cannibalizing
smaller systems that fall within their sphere of influence. The
complete destruction of the victim is usually progressive, and may
take several orbits. However, the stellar debris from the destroyed
dwarf galaxy follows a similar orbital trajectory to the progenitor,
which is likely to have started life far away from the place of its
final demise, and so the tidally disrupted matter tends to be
deposited over a broad range in distance from the larger galaxy. Over
time, with the accumulation of many such mergers, large galaxies
develop an extensive stellar and dark-matter 'halo', the latter being
by far the most massive component of the galaxy. Meanwhile, part of
the (dissipative) gas component of the smaller galaxies feeds the
growth of the disk of the larger galaxy. This is seen in numerical
simulations of galaxy formation, which result in galactic haloes
comprising clumps of dark matter$^5$. If this prediction is correct,
then haloes should possess significant substructure-in contrast to
previous suggestions$^7$, which predict the dark and luminous
components of haloes to be distributed smoothly.

Most of the halo of the Milky Way is metal-poor and, to first order,
smoothly distributed; but recent studies have shown that the halo
contains non-negligible stellar substructure. Evidence for the
phase-space clumping of halo stars has even been found in the solar
neighbourhood, where about 10% of the stars may be associated with a
single ancient accretion event$^4$. The discovery that the Milky Way
is surrounded by a giant rosette-like stream originating from the
Sagittarius dwarf galaxy$^{2,3}$ shows that-on the largest scales-the
structure of the halo is substantially 'streamy': approximately half
of the intermediate-age stars at distances beyond about 15 kpc belong
to the Sagittarius stream. This also implies$^2$ that the last large
accretion was that of the Sagittarius dwarf, and that the Milky Way
has not cannibalized many other small galaxies for about 7 Gyr. Unless
there has been a continual accretion of highly dark-matter-dominated
'galaxies', or of galaxies containing exclusively old stars, the
formation of the Galactic halo must have been essentially complete at
a point in time less than half the age of the Universe.

We now consider whether the behaviour of the Milky Way is unusual. The
only external galaxy in which halo substructure has been reported is
NGC5907, which possesses a gigantic extraplanar stellar stream$^8$,
substantially brighter than similar structures in the Milky
Way. However, this galaxy has long been known to be peculiar$^9$,
having a red, and luminous, flattened 'halo', probably the result of
the strong interaction that deposited the stream. To understand
whether stream-like substructure is the generic morphology of galaxy
haloes, it is necessary to investigate other 'normal' galaxies like
the Milky Way. The prime target for such a study is the Milky Way's
'sister' galaxy, M31-the Andromeda nebula, which at a distance of
$\sim 780$ kpc is the closest large galaxy$^{10}$.

The Wide Field camera$^{11}$ on the 2.5-m Isaac Newton Telescope (INT
WFC) is a four-chip charge-coupled device (CCD) mosaic camera imaging
about 0.3 degree$^2$ per exposure. On the nights of 3-9 September
2000, this instrument was used to survey the southeastern half of the
halo of the Andromeda galaxy. To tile this region out to a distance of
4$^{\circ}$ ($\sim 55$ kpc in projection) from the centre of M31
required 58 contiguous fields. Images were taken in the equivalent of
Johnson visual V and Gunn i bands under good atmospheric conditions,
with 85\% of fields taken in photometric conditions with seeing better
than 1.2 arcsec.

Previous imaging studies of the M31 halo and outer disk have either
sampled the outer parts of M31 at only a few discrete
locations$^{12-14}$, or have taken a panoramic, but much shallower,
view$^{15,16}$. In contrast, our deep survey allows us to make an
uninterrupted study of the spatial variations in stellar density as a
function of magnitude and colour over a large fraction of the
halo. The continuous nature of this survey enables us to distinguish
local density enhancements from both the large-scale structure of the
halo of M31 and the underlying foreground Galactic distribution of
stars. The exposure time of 800-1,000 s per passband per field reaches
i-band magnitude i = 23.5 and V-band magnitude V = 24.5 (with a
signal-to-noise ratio of $\sim 5$) and allows detection of individual
red-giant branch (RGB) stars to an absolute V-band magnitude {\rm $M_V
= 0$} and main-sequence stars to {\rm $M_V = -1$} in the halo of M31.

The WFC pipeline processing provides internal cross-calibration for
the four INT WFC CCDs (each of 4,096 $\times$ 2,048 pixels) at a level
of about 1\% within each pointing. Field-to-field variations in
photometric zero points were calibrated and cross-checked using a
combination of multiple nightly photometric standard sequence
observations and the overlap regions between adjacent WFC
pointings. The overall derived photometric zero-points for the whole
survey are good to the level of 1-2\% in both bands. (Details of the
data processing and calibration will be presented elsewhere.) Objects
were classified as noise artefacts, galaxies or stars according to
their morphological structure on all the images.

Figure 1 presents a summary of the results of our survey. Visual
inspection of the surface density of sources classified as star-like
on the i-band images and with magnitudes and colours consistent with
the known properties of RGB stars at the distance of M31, shows the
presence of a stream-shaped over-density of sources in the halo close
to, but distinct from, the minor axis of M31. The RGB stellar density
in the halo increases on average by a factor of two in the on-stream
regions, and is statistically significant at $\sim$50-sigma. For stars
brighter than the tip of the RGB in M31, the spatial density of
sources is smooth and shows no sign of the feature: the stream
therefore cannot be a foreground Galactic population. The on-stream
stars follow a similar colour-magnitude sequence to the RGB stars in
the remainder of the halo of M31, but with evidence of an enhanced
metallicity relative to the 'normal' M31 halo population owing to
their redder colours. The average V-band surface brightness of the
stream is ${\rm \Sigma_V \approx 30\pm 0.5}$ mag arcsec$^{-2}$, and the stream
extends out to the current limit of our survey, at a projected
distance of about $\sim 40\kpc$.

The metallicity of the stream, deduced from the colour-magnitude
diagrams displayed in Fig. 2, covers a broad range with a mean
slightly more metal-rich than the Galactic globular cluster 47 Tucanae
(whose metallicity, that is, whose ratio of iron to hydrogen compared
to that of the Sun, is {\rm $[Fe/H] = -0.7$}). We note that stars of
near-solar metallicity are concentrated in the stream. Our survey
reveals that such high-metallicity stars are also sparsely distributed
throughout large parts of the halo of M31, consistent with the results
of previous studies$^{12,17}$. This high metallicity, with a mean
approximately 10 times greater than that of the Milky Way's halo, and
the overly large stellar density of the halo-a factor of 10
greater$^{14}$ than that of the Galactic halo-have until now been a
puzzle.

What is this stream-like feature? The outer regions of galactic disks
are metal-poor$^{18}$, so the metallicity of the feature argues
against a direct association with the outer disk of
Andromeda. Furthermore, if it were a disk structure, it would be
located at an impossible de-projected radial distance of about 140 kpc
from the centre of the galaxy. This distance constraint is weakened if
the feature is part of a pronounced disk warp, but that possibility is
highly unlikely given the observed elongated structure. The only
plausible explanation is that it is part of a large stellar stream
within the halo. This stream lies along a line connecting the
Andromeda satellites M32 and NGC205, and is aligned with the direction
of elongation of the outer isophotes of NGC205, suggesting a
relationship between the Andromeda stream and these two dwarf
galaxies. The total absolute magnitude of the stellar stream is {\rm
$M_V \approx -14$}, which we estimate from the differential luminosity
function between equivalent on- and off-stream fields (plausible
assumptions were made to allow for fainter stream stars, the
incompleteness of the survey and the uniformity of the stream). This
luminosity is a factor of about 10 lower than that of either M32 and
NGC205, consistent with the possibility that the stream is the debris
stripped from one (or both) of these two dwarf galaxies during a
recent interaction with M31.

M32 and NGC205 are both unusual dwarf elliptical galaxies which lie at
projected distances of, respectively, only 5 kpc and 9 kpc from the
centre of M31. NGC205 is still active in forming stars, and has had a
varied and complex star-formation history$^{19}$. The central regions are
also known to contain dust, H I gas and molecular gas$^{20}$, and there is
substantial morphological$^{21}$ and kinematic$^{22}$ evidence that indicates
that NGC205 is being tidally distorted and potentially disrupted. The
H I distribution shows a well defined velocity gradient (unlike the
stars), and is less extended than the overall stellar content of the
galaxy$^{20,23}$. The clear inconsistency between the gas and stars
suggests that the gas may have recently been captured, possibly from
the disk of M31.

M32, on the other hand, does not appear to be substantially distorted,
yet it does seem to have a significant population of intermediate-age
stars in addition to a classical old stellar
component$^{24,25}$. There is also clear evidence of a large
metallicity spread in the giant branch, with a mean just below solar,
supporting a prolonged star-forming epoch$^{26}$.

The broad agreement of the metallicity distributions of the stream
stars and these two dwarf satellites, together with their alignment
and physical proximity to M31, point to a common origin. Furthermore,
the unusual properties of most of the rest of the halo of M31 is also
consistent with its relatively recent tidal origin. It seems quite
likely that the apparently peculiar properties of M31's halo are
simply the result of a prolonged, aggressive bout of tidal stripping
from either one, or both, of its two nearest neighbour satellite
galaxies.

If this interpretation is correct, the stream (and possibly most of
the stellar halo) has to be the result of previous interactions with
M31, as there is otherwise not enough time to spatially separate if
from either of the two dwarf galaxies. Furthermore, by comparison with
numerical models of the Sagittarius stream$^{27,28}$, there will be a
similar stream (either leading or trailing the progenitor dwarf(s),
depending on the sense of their orbit) on the opposite side of M31.

During disk-crossing episodes, gas in the disk of M31 and in these
dwarf galaxies is shocked, leading to episodic bursts of star
formation, and gas exchange. In the dwarf galaxies, this gas may
either be recycled from earlier generations of stars or accreted from
the M31 disk. These complex interactions may explain the unusual
star-formation history of the dwarf galaxies and the Andromeda
stream. In turn, the regular impacts perturb the disk of M31, possibly
enough to induce its warped shape$^{29}$ (it is plausible that a similar
process is responsible for the warp in the disk of the Milky Way$^{30}$).

The discovery of the Andromeda stream in the first deep, panoramic
survey of the Milky Way's nearest large companion suggests that halo
substructure in the form of giant tidal streams may be a generic
property of large spiral galaxies, and that the formation of galaxies
continues at a moderate pace up to the present day.  Although the
dwarf galaxies M32 and NGC205 may be the source of this material, we
cannot rule out the alternative possibility that the stream may be the
fossil remnant of a third cannibalized system, found at a stage
intermediate to those seen within the Milky Way$^{2,4}$.

To understand more fully the origin and evolution of the Andromeda
stream, a number of follow-up observational programmes are required,
including an extension of the panoramic survey not only to the
northern regions of the halo of M31, but also out to larger radii, as
it is evident in the present data that the stream extends beyond the
40-kpc limits of the survey. The proximity of M31 also provides us
with an opportunity to undertake a spectroscopic survey of individual
stars within the stream, allowing us to map its kinematic and chemical
properties. As with the tidal material detected in the halo of our own
Galaxy, studies of the Andromeda stream would allow us to map the
distribution of dark matter within the halo of our nearest
neighbouring galaxy, as well as furthering our understanding of the
process of galaxy formation.

\bigskip
\noindent
{\bf REFERENCES}

\def\mnras{\sl Mon. Not. R. astr. Soc.}
\def\apj{\sl Astrophys. J.}
\def\apjl{\sl Astrophys. J.}
\def\aj{\sl Astr. J.}
\def\aap{\sl Astron. \& Astrophys.}
\def\aaps{\sl Astron. \& Astrophys. Suppl.}
\def\araa{\sl Ann. Rev. Astron. \& Astrophys.}
\def\pasp{\sl Pub. Astron. Soc. Pacific}
\def\nat{\sl Nature}

\bibliographystyle{nature}
\bibliography{paper}      

\noindent
  1.  Putman, M. E. et al. Tidal disruption of the Magellanic Clouds
    by the Milky Way. Nature 394, 752-754 (1998). \\
  2.  Ibata, R., Lewis, G. F., Irwin, M., Totten, E. \& Quinn, T. Great
    circle tidal streams: evidence for a nearly spherical massive dark
    halo around the Milky Way. Astrophys. J. 551, 294-311 (2001).\\
  3.  Ibata, R., Irwin, M., Lewis, G. F. \& Stolte, A. Galactic halo
    substructure in the Sloan Digital Sky Survey: The ancient tidal
    stream from the Sagittarius dwarf galaxy. Astrophys. J. 547,
    L133-L136 (2001).\\
  4.  Helmi, A., White, S. D. M., de Zeeuw, P. T. \& Zhao, H. Debris
    streams in the solar neighbourhood as relicts from the formation
    of the Milky Way. Nature 402, 53-55 (1999). \\
  5.  Klypin, A., Gottlöber, S., Kravtsov, A. V. \& Khokhlov,
    A. M. Galaxies in N-body simulations: Overcoming the overmerging
    problem. Astrophys. J. 516, 530-551 (1999).\\
  6.  Cole, S., Aragon-Salamanca, A., Frenk, C. S., Navarro, J. F. \&
    Zepf, S. E. A recipe for galaxy formation. Mon. Not. R.
    Astron. Soc. 271, 781-806 (1994).\\
  7.  Eggen, O. J., Lynden-Bell, D. \& Sandage, A. R. Evidence from the
    motions of old stars that the Galaxy collapsed. Astrophys.
    J. 136, 748-766 (1962).\\
  8.  Shang, Z. et al. Ring structure and warp of NGC 5907:
    Interaction with dwarf galaxies. Astrophys. J. 504, L23-L26
    (1998).\\
  9.  Sackett, P. D., Morrison, H. L., Harding, P. \& Boroson, T. A. A
    faint luminous halo that may trace the dark matter around spiral
    galaxy NGC 5907. Nature 370, 441-443 (1994).\\
 10.  Stanek, K. Z. \& Garnavich, P. M. Distance to M31 with the Hubble
    Space Telescope and HIPPARCOS red clump stars.  Astrophys. J. 503,
    L131-L134 (1998).\\
 11.  Irwin, M. \& Lewis, J. INT WFS pipeline processing. New
    Astron. Rev. 45, 105-110 (2001). \\
 12.  Holland, S., Fahlman, G. G. \& Richer, H. B. Deep HST V- and
    I-band observations of the halo of M31: Evidence for multiple
    stellar populations. Astron. J. 112, 1035-1045 (1996). \\
 13.  Rich, R. M., Mighell, K. J., Freedman, W. L. \& Neill,
    J. D. Local Group populations with the Hubble Space
    Telescope. I. The M31 globular cluster G1=Mayall
    II. Astron. J. 111, 768-776 (1996). \\
 14.  Reitzel, D. B., Guhathakurta, P. \& Gould, A. Isolating red giant
    stars in M31's elusive outer spheroid. Astron. J. 116, 707-722
    (1998). \\
 15.  Walterbos, R. A. M. \& Kennicutt, R. C. Multi-color photographic
    surface photometry of the Andromeda galaxy. Astron.
    Astrophys. Suppl. 69, 311-332 (1987).\\
 16.  Innanen, K. A., Kamper, K. W., van den Bergh, S. \& Papp,
    K. A. The optical warp of M31. Astrophys. J. 254, 515-516 (1982).\\
 17.  Durrell, P. R., Harris, W. E., Pritchet, C. J. \& Davidge,
    T. Photometry of the outer halo of M31. Am. Astron. Soc. Meeting
    195, 4.04 (1999).\\
 18.  Ferguson, A. M. N., Gallagher, J. S. \& Wyse, R. F. G. The
    extreme outer regions of disk galaxies. I. Chemical abundances of
    H II regions. Astron. J. 116, 673-690 (1998). \\
 19.  Lee, M. G. Stellar population is the central region of the dwarf
    elliptical galaxy NGC 205. Astron. J. 112, 1438-1449 (1996). \\
 20.  Young, L. M. \& Lo, K. Y. The neutral interstellar medium in
    nearby dwarf galaxies. II. NGC 185, NGC 205, and NGC 147.
    Astrophys. J. 476, 127-143 (1997).\\
 21.  Hodge, P. W. The structure and content of NGC
    205. Astrophys. J. 182, 671-696 (1973).\\
 22.  Bender, R., Paquet, A. \& Nieto, J. Internal stellar kinematics
    of three dwarf ellipticals in the Local Group. Astron. Astrophys.
    246, 349-353 (1991).\\
 23.  Welch, G. A., Sage, L. J. \& Mitchell, G. F. The puzzling
    features of the interstellar medium in NGC 205. Astrophys. J. 499,
    209-220 (1998).\\
 24.  Freedman, W. L. Stellar content of nearby galaxies. II--the
    Local Group dwarf elliptical galaxy M32. Astron. J. 98, 1285-1304
    (1989). \\
 25.  Davidge, T. J. The evolved red stellar content of
    M32. Publ. Astron. Soc. Pacif. 112, 1177-1187 (2000).\\
 26.  Grillmair, C. J. et al. Hubble Space Telescope observations of
    M32: The color-magnitude diagram. Astron. J. 112, 1975-1987
    (1996). \\
 27.  Johnston, K. V., Spergel, D. N. \& Hernquist, L. The disruption
    of the Sagittarius dwarf galaxy. Astrophys. J. 451, 598-606
    (1995).\\
 28.  Ibata, R. A. \& Lewis, G. F. Galactic indigestion: Numerical
    simulations of the Milky Way's closest
    neighbor. Astrophys. J. 500, 575-590 (1998).\\
29.  Schwarzschild, M. Mass distribution and mass-luminosity ratio in
    galaxies. Astron. J. 59, 273-284 (1954). \\
30.  Ibata, R. A. \& Razoumov, A. O. Archer of the Galactic disk? The
    effect on the outer HI disk of the Milky Way of collisional
    encounters with the Sagittarius dwarf
    galaxy. Astron. Astrophys. 336, 130-136 (1998).\\

\bigskip

\noindent
{\bf Acknowledgements.} This paper is based on observations made with the
Isaac Newton Telescope operated on the island of La Palma by the Isaac
Newton Group in the Spanish Observatorio del Roque de los Muchachos of
the Instituto de Astrofisica de Canarias.

\newpage

\bibliographystyle{nature}
\bibliography{paper}      

\newpage
\epsfxsize=12.0cm \epsfbox[38 104 575 690]{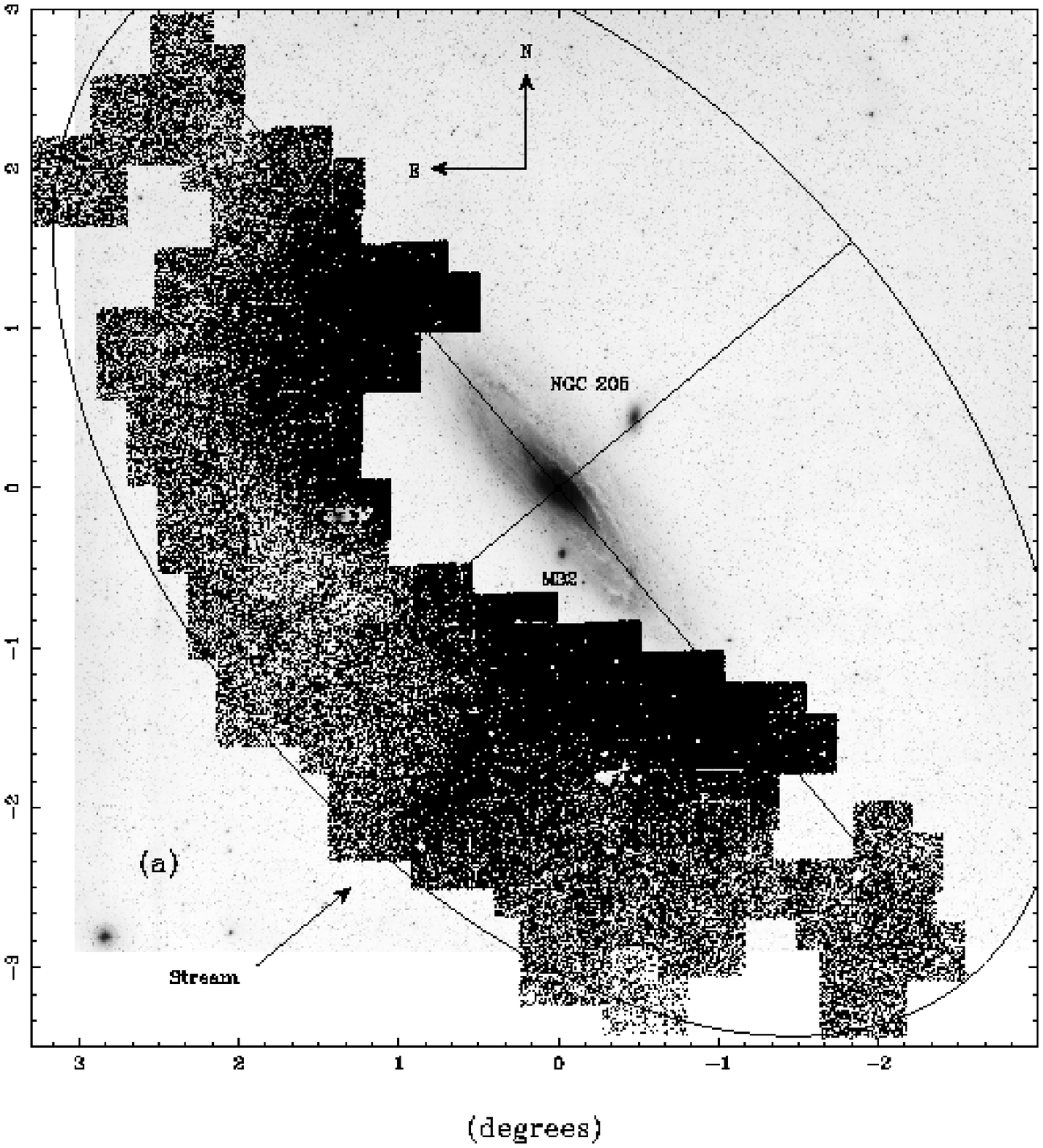}

\noindent
{\bf Figure 1.} Surface density of red-giant branch (RGB) stars over
the outer southeastern halo of M31. The surface-density maps produced
by our survey are shown as 'tiles' overlying an optical image centred
on M31 (see below). The over-density of stars revealed in the present
study is seen as a stream extending out of M31 close to its minor
axis. The density of extended background galaxies (approximately
25,000 per degree$^2$ to $V < 24$ mag) becomes comparable to the
stellar density at a projected distance of 20-25 kpc from the centre
of M31 along the minor axis. A fraction of these galaxies are
misclassified by the reduction algorithm, thereby contaminating the
stellar sample. There is a further contamination from foreground
Galactic disk dwarfs, which number $<10$\% of the M31 halo population
over the survey region. However, the contribution of these
contaminating sources to the number density maps is easily
removed. Indeed, the stream can be detected at signal-to-noise ratio
$>5$ over tiny individual regions of area 0.01 degree$^2$. The gaps to
the northeast and southwest of the map correspond to fields taken in
poor seeing conditions where the image quality criteria of the survey
were not attained. The surface density maps are superimposed upon a
photographic Palomar Sky Survey image of M31; clearly visible are the
two companion galaxies M32 and NGC 205. The major and minor axes of
M31 are displayed, and an ellipse denoting the size of a flattened
ellipsoid (aspect ratio 3:5) of semi-major axis length 55 kpc has been
superimposed to show the spatial extent of the survey. The location of
the Andromeda stream is labelled. The overall shape of the M31 halo
revealed by our observations appears rather boxy, with indications of
other possible substructures-but these results require further study
to confirm their reality.

\newpage
\hskip -1cm
\epsfysize=22cm \epsfbox{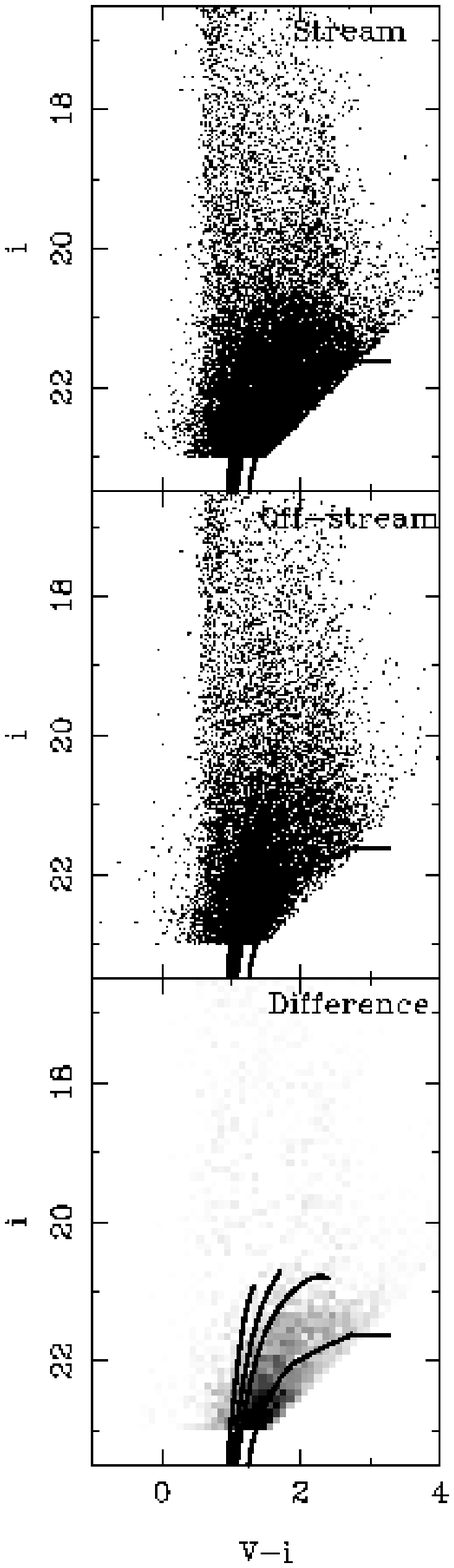}

\vskip -22cm
\hskip 5.75cm
\vbox{
\hsize=8.5cm
\noindent
{\bf Figure 2.} The relation between V-i colour and i-band
magnitude. Data are shown for a stream field (a), and an off-stream
field (b). a and b both cover 0.3 degree$^2$ of sky, and were taken in
similar observing conditions (0.97 arcsec in i and 0.93 arcsec in V
for the stream field, and 1.08 arcsec for both i and V for the
off-stream field). a and b are also at a similar distance
($\sim 2^{\circ}$) from the centre of M31, with the off-stream field almost
exactly on the minor axis of the galaxy. The red-giant branch (RGB)
stars of M31 are seen at magnitudes fainter than $i = 20.5$. To guide
the interpretation of these diagrams, we have superimposed on each
panel the RGB tracks (shifted to the distance of M31 and corrected for
extinction) of four well-studied galactic globular clusters of
different metallicity ([Fe/H]); these are, from left to right, NGC6397
{\rm $([Fe/H] = -1.91)$}, NGC1851 {\rm $([Fe/H] = -1.29)$}, 47 Tuc
{\rm $([Fe/H] = -0.71)$} and NGC6553 {\rm $([Fe/H] = -0.2)$}. The
difference between the stellar populations in a and b is displayed in
c in the form of a binned colour-magnitude diagram. With much of the
contamination removed in this way, we see that the stream contains a
broad range of stellar populations, from metallicities similar to that
of NGC1851 through to solar abundance, with a mean metallicity
slightly higher than that of 47 Tuc. It is particularly difficult to
constrain the distance to the stream owing to this metallicity spread;
to within an uncertainty of possibly as much as 0.5 mag the stream RGB
appears at the same position as the M31 RGB, which suggests a distance
of $d\approx 800 \pm 200$ kpc.  (The photometric uncertainties are
approximately 0.2 mag at the faint limit of the data in the diagrams;
0.1 mag uncertainties occur at $V = 23.7$ mag, $i = 22.5$ mag). }

\end{document}